\documentclass[10pt]{article}
\usepackage{graphicx}

\def\Title#1{\begin{center} {\Large #1 } \end{center}}
\def\Author#1{\begin{center}{ \sc #1} \end{center}}
\def\Address#1{\begin{center}{ \it #1} \end{center}}

\newcommand\pubblock{\rightline{\begin{tabular}{l} Proceedings of the Fifth Annual LHCP\\ \pubnumber\\
         \pubdate  \end{tabular}}}

\newenvironment{Abstract}{\begin{quotation} \begin{center} 
             \large ABSTRACT \end{center}\bigskip 
      \begin{center}\begin{large}}{\end{large}\end{center} \end{quotation}}

\newenvironment{Presented}{\begin{quotation} \begin{center} 
             PRESENTED AT\end{center}\bigskip 
      \begin{center}\begin{large}}{\end{large}\end{center} \end{quotation}}

\def\beq{\begin{equation}}
\def\eeq#1{\label{#1}\end{equation}}
\def\eeqn{\end{equation}}

\def\beqa{\begin{eqnarray}}
\def\eeqa#1{\label{#1}\end{eqnarray}}
\def\eeqan{\end{eqnarray}}

\let\bar=\overbar

\def\Dslash{\not{\hbox{\kern-4pt $D$}}}
\def\dslash{\not{\hbox{\kern-2pt $\del$}}}

\def\msb{{\bar{\ssstyle M \kern -1pt S}}}

\usepackage{color}

\newcommand\pubnumber{ CMS-CR-2017-223 }

\newcommand\pubdate{\today}

\def\affiliation{
on behalf of the CMS Experiment, \\
Vinca Institute of Nuclear Sciences \\
Belgrade University, Belgrade, 11001, Serbia}

\begin{document}

\large
\begin{titlepage}
\pubblock

\vfill
\Title{  NEW RESULTS ON COLLECTIVITY WITH CMS  }
\vfill

\Author{ JOVAN MILOSEVIC  }
\Address{\affiliation}
\vfill
\begin{Abstract}

Nonlinear response coefficients of higher-order anisotropy harmonics for charged particles are measured in PbPb collisions at 2.76 and 5.02 TeV as a function of transverse momentum and collision centrality. The nonlinear response coefficients are obtained from $v_{n}$ harmonics measured with respect to their own plane and the mixed harmonics. Additionally, at 5.02 TeV PbPb collisions, a significant negative skewness is observed by a fine splitting between $v_{2}\{4\}$ and $v_{2}\{6\}$ cumulants. The elliptic flow skewness is measured up to 60\% centrality. The results are compared with hydrodynamic models with different shear viscosity to entropy density ratios and initial conditions.

\end{Abstract}
\vfill

\begin{Presented}
The Fifth Annual Conference\\
 on Large Hadron Collider Physics \\
Shanghai Jiao Tong University, Shanghai, China\\ 
May 15-20, 2017
\end{Presented}
\vfill
\end{titlepage}
\def\thefootnote{\fnsymbol{footnote}}
\setcounter{footnote}{0}
%

\normalsize 


\section{Introduction}

Investigation of hydrodynamical behavior of Quark Gluon Plasma (QGP) through measurements of $v_{n}$ harmonics indicates that higher order harmonics ($n > $3) have a non-linear contribution~\cite{Niemi:2012aj} and may be composed of lower order harmonics ($n < $4)~\cite{Yan:2015jma}. A deeper insight into the influence of the initial-state fluctuations onto the hydrodynamic flow can be obtained by studying the correlations between higher order harmonics $v_{n}\{\Psi_{n}\}$ measured with respect to their own plane, and mixed higher order harmonics $v_{n}\{\Psi_{mkl}\}$ ($m,k,l < n$) measured with respect to the direction of multiple lower order harmonics. In Ref.~\cite{Yan:2015jma,Qian:2016fpi} it was shown that, to a good approximation, the following relations decompose higher order harmonics into their linear ($V_{nL}$) and non-linear contributions.
\begin{eqnarray}
\label{sum}
V_{4}&=&V_{4L}+\chi_{422}(V_{2})^{2} \nonumber \\
V_{5}&=&V_{5L}+\chi_{523}V_{2}V_{3} \nonumber \\ 
V_{6}&=&V_{6L}+\chi_{6222}(V_{2})^{3}+\chi_{633}(V_{3})^{2} \nonumber \\
V_{7}&=&V_{7L}+\chi_{723}(V_{2})^{2}V_{3}
\end{eqnarray}
The corresponding definitions of the mixed higher order harmonics $v_{4}\{\Psi_{22}\}$, $v_{5}\{\Psi_{23}\}$, $v_{6}\{\Psi_{222}\}$, $v_{6}\{\Psi_{33}\}$ and $v_{7}\{\Psi_{223}\}$, and the non-linear response coefficients $\chi_{422}$, $\chi_{523}$, $\chi_{6222}$, $\chi_{633}$ and $\chi_{723}$ can be found in~Ref.~\cite{Yan:2015jma,Qian:2016fpi}. As the mixed harmonics and the nonlinear response coefficients are sensitive to the initial-state conditions and transport properties of the formed medium, they could be used to test theoretical models and to make distinction between them.

Another observable which could make constraints on the initial-state conditions is the skewness~\cite{Yan:2014afa}. As a consequence of the non-gaussian fluctuations of the participant eccentricity a fine splitting between $v_{2}\{4\}$ and $v_{2}\{6\}$ magnitudes obtained from 4- and 6-particle cumulants appears. Then the skewness which is experimentally defined as
\begin{equation}
\label{skew}
\gamma^{exp}_{1} = -6\sqrt{2}v_{2}\{4\}^{2}\frac{v_{2}\{4\} - v_{2}\{6\}}{(v_{2}\{2\}^{2} - v_{2}\{4\}^{2})^{3/2}}
\end{equation}
can be measured and compared with theoretical models.
\section{Experiment and data used}
The CMS tracker detector is surrounded with a super-conducting solenoid producing $\mathrm{3.8}$~T magnetic field which enable precise $p_{T}$ measurements above 0.3 GeV/c with a typical resolution of 1.5\% in $\mathrm{p_{T}}$. About 30 milion minimum-bias PbPb collisions at the LHC energies of $\mathrm{\sqrt{s_{NN}} = 2.76}$~TeV and 100 milion at $\mathrm{5.02}$~TeV were collected. The results of these analyses are based on the silicon tracker and hadron forward (HF) calorimeters. The tracks with $p_{T} > $0.3~GeV/c are used. A wide pseudorapidity coverage of $\mathrm{|\eta| < }$~2.5 for the tracker and 2.9~$\mathrm{ < |\eta| < }$~5.2 for the HF calorimeters together with a full azimuthal coverage excellently suites for studying the long-range correlations. More detailed description of the CMS detector can be found in Ref.~\cite{Chatrchyan:2008aa}.
\section{Results}
\begin{figure}[htb]
\centering
\includegraphics[height=2.55in]{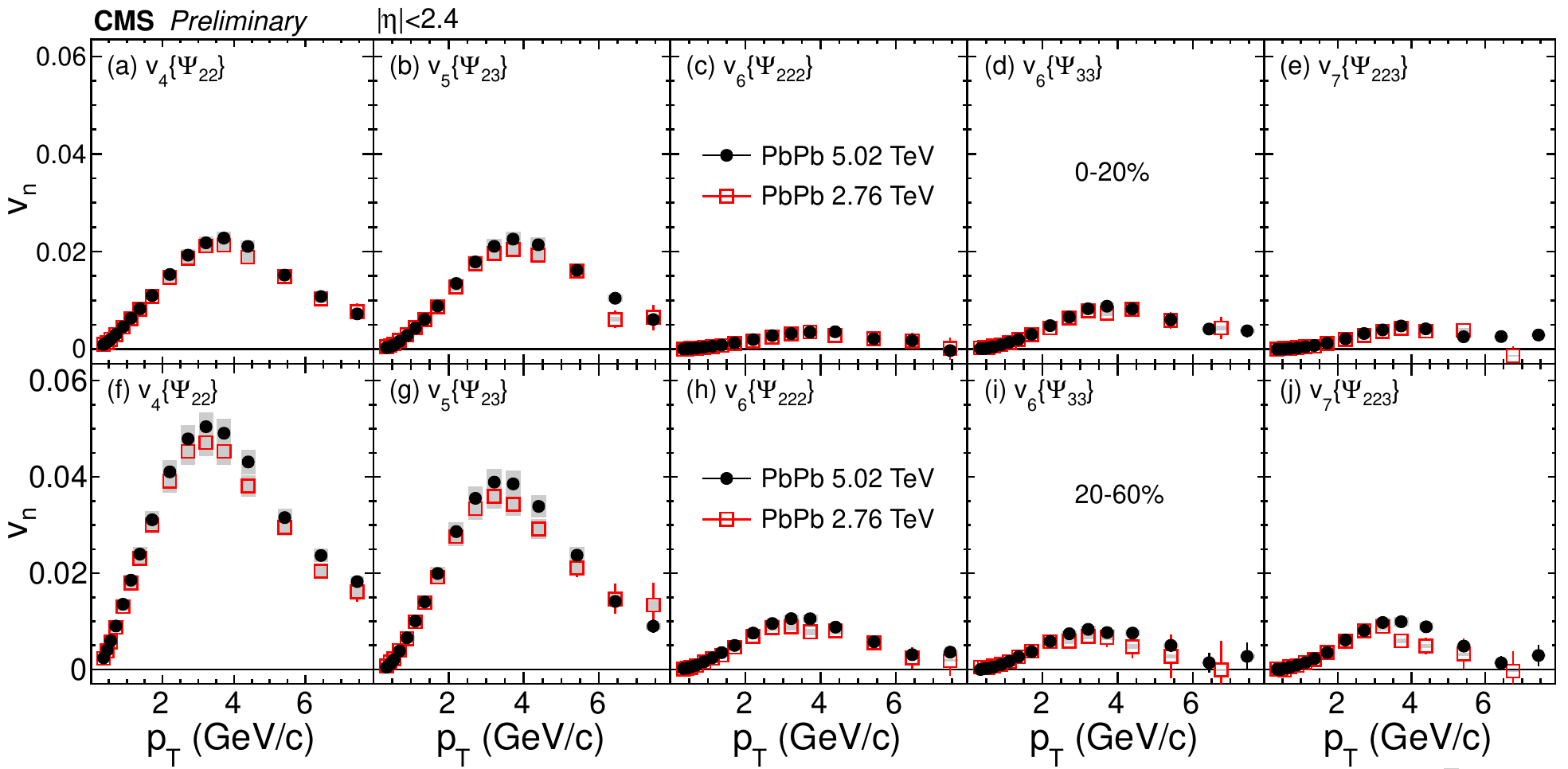}
\caption{The mixed higher order harmonics {\it vs} $p_{T}$ measured using the SP method in PbPb collisions at 2.76 and 5.02~TeV~\cite{CMS:2017ltu}. Top row corresponds to 0--20\% and bottom row to 20--60\% centrality range. Statistical (systematic) uncertainties are shown as error bars (shadow boxes).}
\label{fig:figure1}
\end{figure}

In Fig.~\ref{fig:figure1} are shown mixed higher order harmonics as a function of $p_{T}$ measured using the Scalar Product (SP) method in PbPb collisions at 2.76 and 5.02 TeV~\cite{CMS:2017ltu}. The measurement is performed in two centrality classes: 0--20\% and 20--60\%. The $v_{5}\{\Psi_{23}\}$, $v_{6}\{\Psi_{33}\}$ and $v_{7}\{\Psi_{223}\}$ are measured for the first time. Mixed harmonics of all measured orders show a very weak energy dependence. As expected, their magnitudes increase going from central to peripheral collisions.

\begin{figure}[h!]
\centering
\includegraphics[height=2.55in]{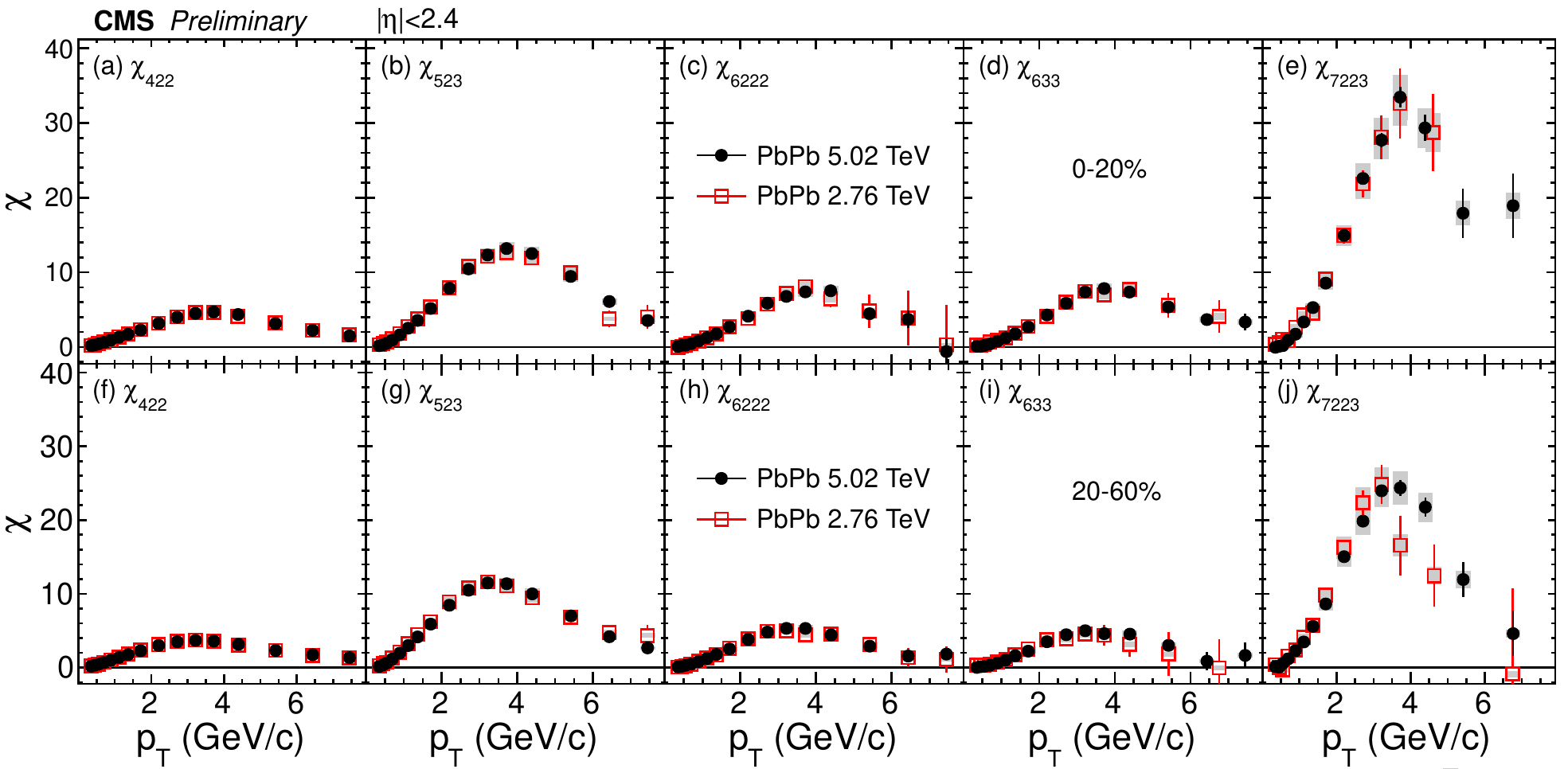}
\caption{The non-linear response coefficients $\chi$ {\it vs} $p_{T}$ measured using the SP method in PbPb collisions at 2.76 and 5.02~TeV~\cite{CMS:2017ltu}. Top row corresponds to 0--20\% and bottom row to 20--60\% centrality range. Statistical (systematic) uncertainties are shown as error bars (shadow boxes).}
\label{fig:figure3}
\end{figure}

Figure \ref{fig:figure3} shows non-linear response coefficients $\chi$ as a function of $p_{T}$ in PbPb collisions at 2.76 and 5.02~TeV for the same centrality classes as those from Fig.~\ref{fig:figure1}. The non-linear response coefficients $\chi_{422}$, $\chi_{523}$, $\chi_{6222}$, $\chi_{633}$ and $\chi_{723}$ are measured for the first time. The odd, $\chi_{523}$ and $\chi_{723}$, have a stronger non-linear response with respect to the even harmonics. For two analyzed energies, there is nearly no energy and a weak centrality dependence of the non-linear response coefficients.

\begin{figure}[htb]
\centering
\includegraphics[height=1.5in]{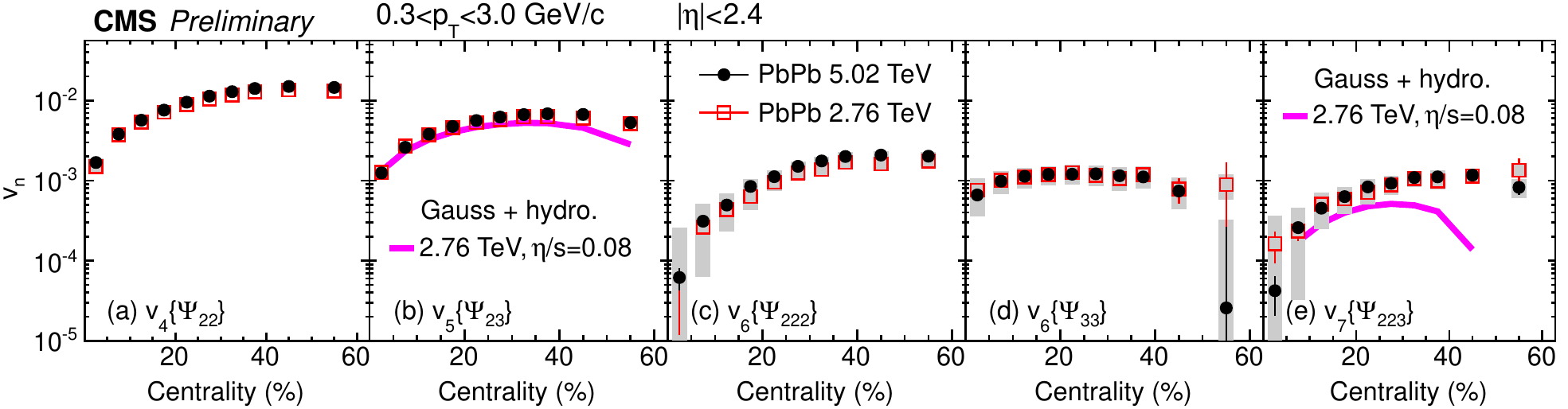}
\caption{The mixed higher order harmonics {\it vs} centrality in PbPb collisions at 2.76 and 5.02~TeV~\cite{CMS:2017ltu}. Statistical (systematic) uncertainties are shown as error bars (shadow boxes). The line represents a hydrodynamic calculation performed with a deformed symmetric Gaussian density profile~\cite{Yan:2015jma} with $\eta/s = 0.08$ in 2.76 TeV PbPb collisions.}
\label{fig:figure4}
\end{figure}

Centrality dependence of mixed order harmonics, averaged over $0.3 < p_{T} < 3.0$~GeV/c, for the two energies in PbPb collisions is presented in Fig.~\ref{fig:figure4}. Again, a weak energy dependence is seen. Except for $v_{6}\{\Psi_{33}\}$, mixed harmonics exhibit a strong centrality dependence. In the same figure is drawn a hydrodynamic calculations with a deformed symmetric Gaussian density profile~\cite{Yan:2015jma} for $v_{5}\{\Psi_{23}\}$ and for $v_{7}\{\Psi_{223}\}$ calculated with $\eta/s = 0.08$ for PbPb collisions at 2.76 TeV. The calculations describes rather well $v_{5}\{\Psi_{23}\}$ data, but not $v_{7}\{\Psi_{223}\}$.

\begin{figure}[htb]
\centering
\includegraphics[height=1.5in]{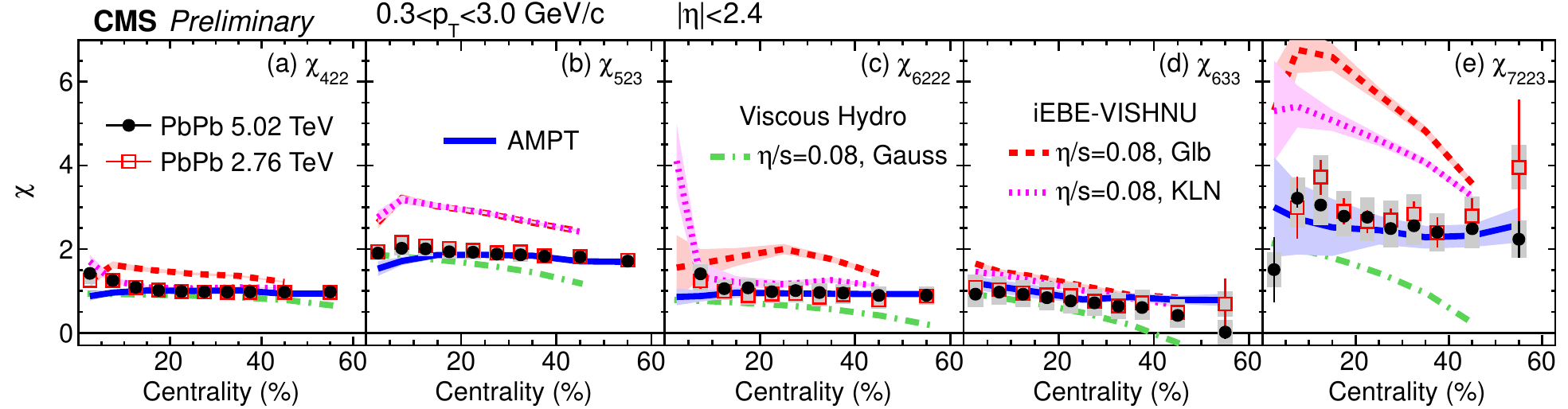}
\caption{The non-linear response coefficients {\it vs} centrality in PbPb collisions at 2.76 and 5.02~TeV~\cite{CMS:2017ltu}. Statistical (systematic) uncertainties are shown as error bars (shadow boxes). The lines represent different hydrodynamic models~\cite{Yan:2015jma,Qian:2016fpi} with $\eta/s = 0.08$ at different initial-state conditions. The full blue line is the calculation from AMPT model.}
\label{fig:figure5}
\end{figure}

Fig.~\ref{fig:figure5} shows the centrality dependence, again averaged over $0.3 < p_{T} < 3.0$~GeV/c, for the corresponding non-linear response coefficients. In contrast to the mixed harmonics results, the non-linear response coefficients do not show a strong energy and centrality dependence. The experimental data for all harmonics are described well by AMPT model. Comparisons to hydrodynamical model with a deformed symmetric Gaussian density profile~\cite{Yan:2015jma} and with iEBE-VISHNU  model~\cite{Qian:2016fpi}, where both calculations have been performed with $\eta/s = 0.08$, show a strong sensitivity to the initial-state conditions. The model results also shows that the sensitivity of the non-linear response coefficients increases with an increase of the harmonic order n.

\begin{figure}[h!]
\centering
\includegraphics[height=1.5in]{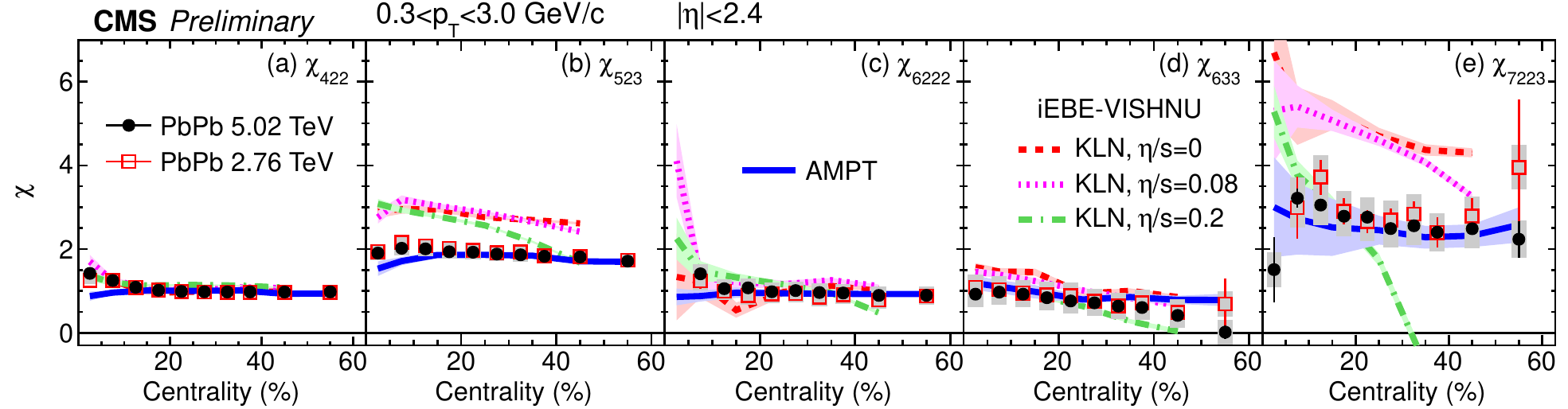}
\caption{The non-linear response coefficients {\it vs} centrality in PbPb collisions at 2.76 and 5.02~TeV~\cite{CMS:2017ltu}. Statistical (systematic) uncertainties are shown as error bars (shadow boxes). The lines represent hydrodynamic results~\cite{Qian:2016fpi} obtained at a given KLN initial-state condition under different $\eta/s$ values. The full blue line is the calculation from AMPT model.}
\label{fig:figure6}
\end{figure}

In Fig.~\ref{fig:figure6} are shown the same experimental data as in Fig.~\ref{fig:figure5}, compared with iEBE-VISHNU model~\cite{Qian:2016fpi}, computed with the same KLN initial-state condition but with different $\eta/s$ values. The non-linear response coefficients are sensitive to the $\eta/s$ values, especially for the odd harmonics.

\begin{figure}[htb]
\centering
\includegraphics[height=2.00in]{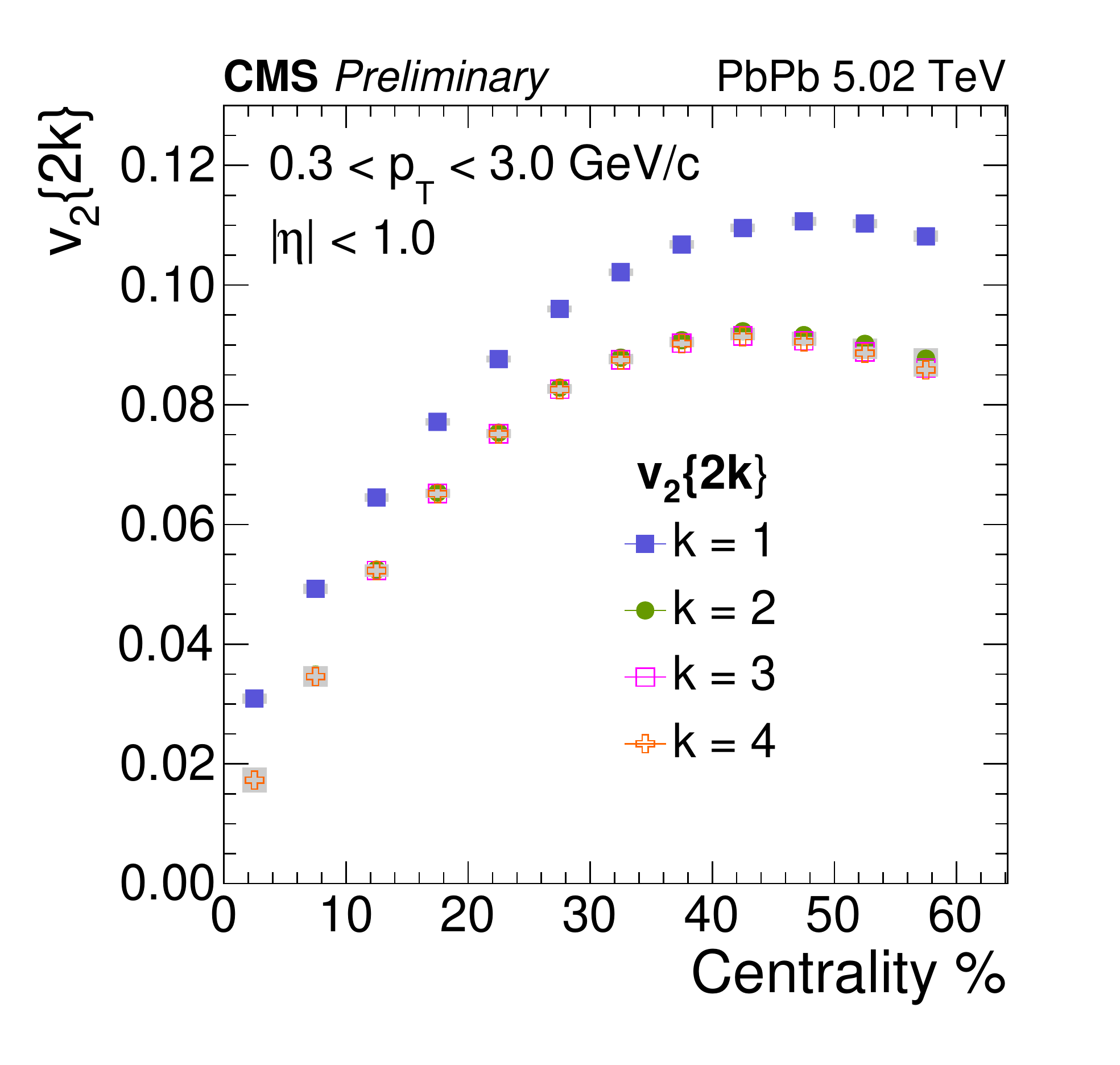}
\caption{Elliptic flow cumulant results as a function of centrality~\cite{CMS:2017zgs}. Statistical uncertainties are covered by the symbol size. Systematic uncertainties are shown as gray bands.}
\label{fig:figure7}
\end{figure}

\begin{figure}[htb]
\centering
\includegraphics[height=2.00in]{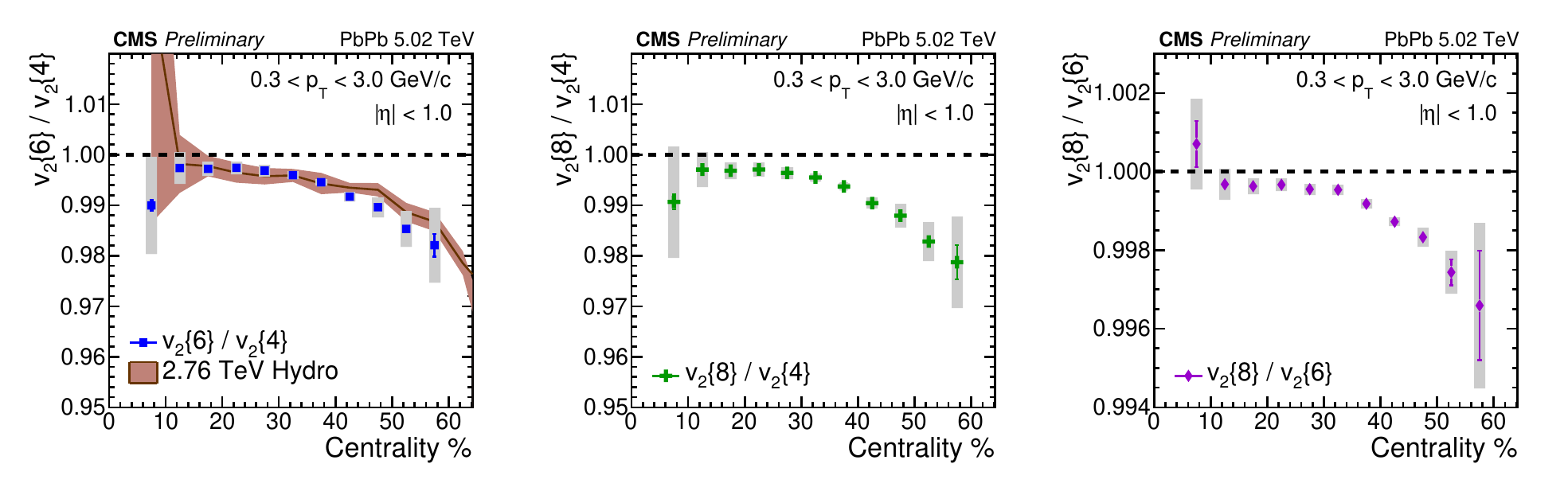}
\caption{Ratios of higher-order cumulant elliptic flow harmonics~\cite{CMS:2017zgs}. Both statistical (error bars) and systematic (gray bands) uncertainties are shown. Hydrodynamic result for 2.76 TeV PbPb collisions from Ref.~\cite{Giacalone:2016eyu} is presented as a colored band and are compared to the measured $v_{2}\{6\}/v_{2}\{4\}$ ratio.}
\label{fig:figure8}
\end{figure}

In Fig.~\ref{fig:figure7} are depicted results for the centrality dependence of the elliptic flow extracted using cumulants of different orders $v_{2}\{2k\}$ \cite{CMS:2017zgs}. Roughly, the $v_{2}\{2k\}$ results show an expected ordering $v_{2}\{2\} > v_{2}\{4\} \approx v_{2}\{6\} \approx v_{2}\{8\}$. A weak splitting of the higher-order cumulants, especially in peripheral collisions, is visible. The splitting can be seen better by looking at the ratios between the higher-order cumulants. These ratios are shown in Fig.~\ref{fig:figure8} in which one can clearly see that the earlier observation $v_{2}\{4\} \approx v_{2}\{6\} \approx v_{2}\{8\}$ is not fully valid. Namely, a fine splitting between higher-order cumulants exists, and they are ordered as $v_{2}\{4\} > v_{2}\{6\} > v_{2}\{8\}$. The effect is small, on the percent level. The magnitude of the effect increases going from central to peripheral collisions. The event-by-event hydrodynamic result with $\eta/s = $0.08 for PbPb collisions at 2.76 TeV, taken from Ref.~\cite{Giacalone:2016eyu}, is consistent with the experimental measurement at 5.02 TeV. As any changes in the initial state eccentricities between 2.76 and 5.02 TeV are expected to be small, the similarity between the experimental results for the two energies is expected.

\begin{figure}[htb]
\centering
\includegraphics[height=2.00in]{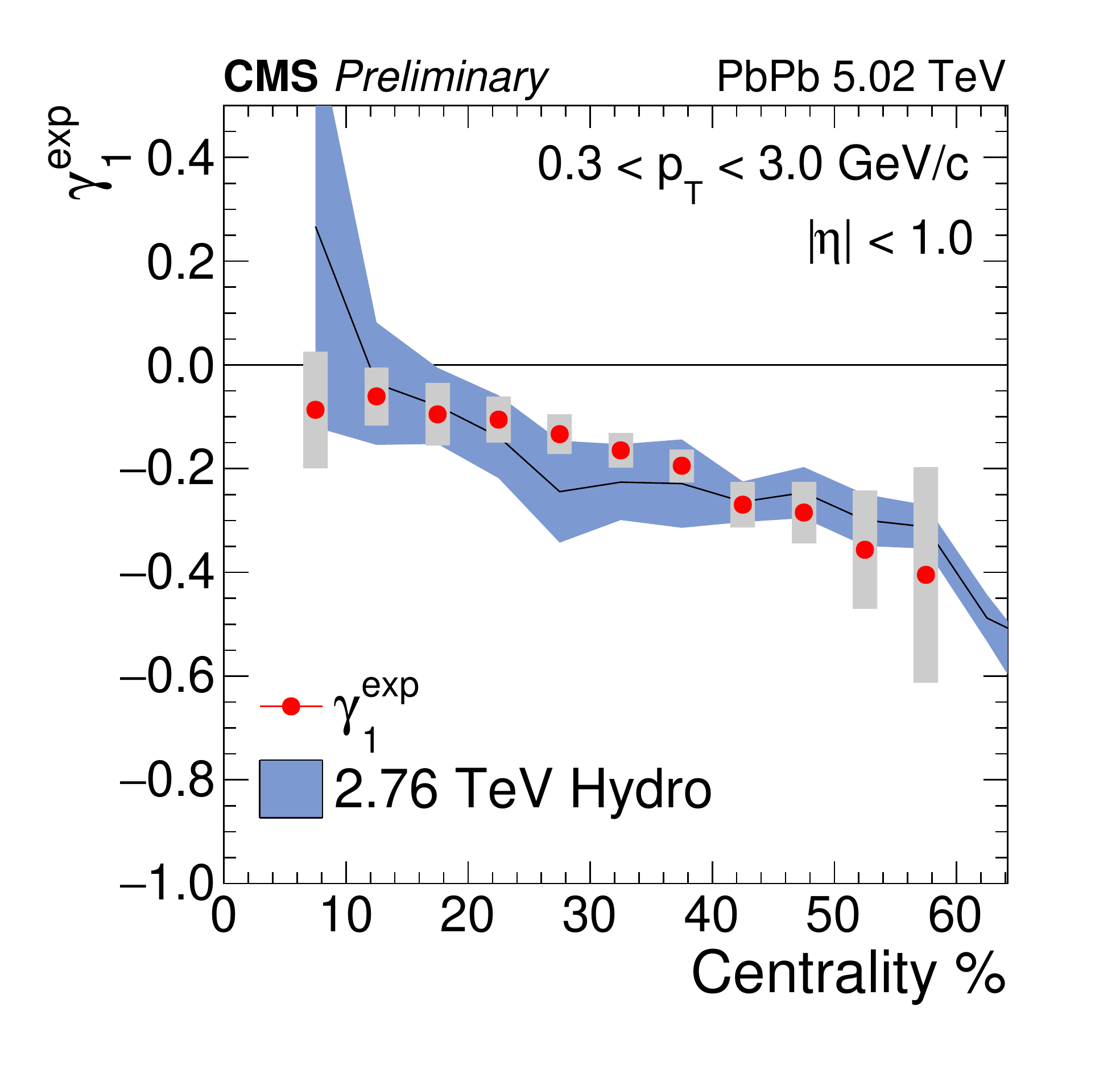}
\caption{The skewness calculated according to the Eq.(\ref{skew}) in PbPb collisions at 5.02 TeV~\cite{CMS:2017zgs}. Statistical and systematic uncertainties are shown. Hydrodynamic result for 2.76 TeV PbPb collisions from Ref.~\cite{Giacalone:2016eyu} is shown as a colored band.}
\label{fig:figure10}
\end{figure}

Fig.~\ref{fig:figure10} shows the skewness, $\gamma^{exp}_{1}$, for PbPb collisions at 5.02 TeV calculated according to the Eq.(\ref{skew}) and plotted {\it vs} centrality~\cite{CMS:2017zgs}. It has non-zero values and the size of the effect increases going from central to peripheral collisions. These experimental results are in a very good agreement with the hydrodynamic result for the skewness in PbPb collisions at 2.76 TeV, taken from Ref.~\cite{Giacalone:2016eyu}.

\section{Summary}
The CMS Collaboration has been measured for the first time the mixed higher order $v_{n}$ harmonics and nonlinear response coefficients of charged particles as a function of $p_{T}$ and centrality in PbPb collisions at 2.76 TeV and 5.02 TeV. The nonlinear contribution for the odd harmonics are larger than for the even ones, especially in the centrality range 20-60\%. The data are compared with AMPT and hydrodynamic models with different $\eta/s$ values and initial conditions. The AMPT model is favored by the measurement. The results will provide constraints on the theoretical description of the medium close to the freeze-out temperature, which is poorly understood so far.

A non-Gaussian behavior is observed for the event-by-event fluctuations of the $v_{2}$ coefficients in PbPb collisions at 5.02 TeV. It manifests as splitting of higher order cumulants which are ordered as $v_{2}\{4\} > v_{2}\{6\} > v_{2}\{8\}$ for non-central events with centralities greater than $\approx$ 15\%. The skewness, $\gamma^{exp}_{1}$, quantify the effect, and is found to be negative with an increasing magnitude as collisions become less central. Presented results will provide constraints on the theoretical description of the medium formed in heavy-ion collisions.


\begin{thebibliography}{99}


\bibitem{Niemi:2012aj} 
  H. Niemi, and G. S. Denicol, and H. Holopainen, and P. Huovinen,
  Phys.\ Rev.\ C {\bf 87}, 054901 (2013)
  [arXiv:1212.1008 [nucl-th]].

\bibitem{Yan:2015jma} 
  L. Yan, and J.-Y. Ollitrault,
  Phys.\ Lett.\ B {\bf 744}, 82 (2015)
  [arXiv:1502.02502 [nucl-th]].

\bibitem{Qian:2016fpi} 
  J. Qian, and U. W. Heinz, and J. Liu,
  Phys.\ Rev.\ C {\bf 93}, 064901 (2016)
  [arXiv:1602.02813 [nucl-th]].

\bibitem{Yan:2014afa} 
  L. Yan, and J.-Y. Ollitrault, and A. M. Poskanzer,
  Phys.\ Rev.\ C {\bf 90}, 024903 (2014)
  [arXiv:1405.6595 [nucl-th]].

\bibitem{Chatrchyan:2008aa}
   CMS Collaboration,
   JINST {\bf 3}, S08004 (2008)

\bibitem{CMS:2017ltu}
  CMS Collaboration,
  CMS-PAS-HIN-16-018

\bibitem{CMS:2017zgs}
  CMS Collaboration,
  CMS-PAS-HIN-16-019


\bibitem{Giacalone:2016eyu} 
  G. Giacalone, and L. Yan, and J. Noronha-Hostler, and J.-Y. Ollitrault,
  Phys.\ Rev.\ C {\bf 95}, 014913 (2017)
  [arXiv:1608.01823 [nucl-th]].


\end{thebibliography}
\end{document}